\documentclass{JHEP3}
\usepackage{amsmath, amsfonts}
\usepackage{epsfig}
\usepackage{cite}

\newcommand{\ba}{\begin{eqnarray}}
\newcommand{\ea}{\end{eqnarray}}
\newcommand{\rmi}[1]{{\mbox{\scriptsize #1}}}
\newcommand{\rmii}[1]{{\mbox{\tiny #1}}}

\newcommand{\tr}{{\rm Tr\,}}
\newcommand{\nn}{\nonumber \\}
\newcommand{\fr}[2]{{\frac{#1}{#2}\,}}
\newcommand{\msbar}{{\overline{\mbox{\rm MS}}}}

\renewcommand{\vec}[1]{{\bf #1}}

\renewcommand{\(}{\left(}
\renewcommand{\)}{\right)}

\newcommand{\MB}{{\bar{\mu}_{\rm{3d}}}}

\renewcommand{\ln}{{\rm ln}}

\def\half{{\textstyle \frac 12}}

\def\gE{g_3}

\def\x{\mathbf {x}}
\def\Z{\mathcal{Z}}
\def\zp{\bar z}
\def\vv{\bar v}
\def\openone{\rlap 1\kern 0.22ex 1}

\def\gauge{s}

\newcommand{\heavy}{\Sigma}
\newcommand{\light}{\Pi}
\newcommand{\heavys}{\phi}
\newcommand{\lights}{\chi}
\newcommand{\heavye}{\rho}
\newcommand{\lighte}{\omega}

\preprint{arXiv.org/0801.1566 [hep-ph]\\ CERN-PH-TH/2008-007 \\ HIP-2008-01/TH \\ TUW-08-01 }

\title
    {%
    Center-Symmetric Effective Theory for High-Temperature SU(2) Yang-Mills Theory
     }

\author
    {%
    Ph.~de Forcrand$^{12}$\footnote{\tt forcrand@phys.ethz.ch}, A.~Kurkela$^3$\footnote{\tt aleksi.kurkela@helsinki.fi}~ and A.~Vuorinen$^{24}$\footnote{\tt vuorinen@phys.washington.edu}
    \\
   $^1$Institut f\"ur Theoretische Physik, ETH Z\"urich, CH-8093 Z\"urich, Switzerland\\
   $^2$CERN, Physics Department, TH Unit, CH-1211 Geneva 23, Switzerland\\
   $^3$Theoretical Physics Division, Department of Physical Sciences, P.O.~Box 64, FI-00014 University of Helsinki, Finland\\
   $^4$Institut f\"ur Theoretische Physik, Technische Universit\"at Wien, Wiedner Hauptstr. 8-10, A-1040 Vienna, Austria

    }

\abstract{
We construct and study a dimensionally reduced effective theory for high-temperature SU(2) Yang-Mills theory that respects all the symmetries of the underlying theory. Our main motivation is to study whether the correct treatment of the center symmetry can help extend the applicability of the dimensional reduction procedure towards the confinement transition. After performing perturbative matching to the full theory at asymptotically high temperatures, we map the phase diagram of the effective theory using non-perturbative lattice simulations. We find that at lower temperature the theory undergoes a second order confining phase transition, in complete analogy with the full theory, which is a direct consequence of having incorporated the center symmetry.
}

\keywords{Thermal Field Theory, QCD, Lattice Field Theory}

\begin{document}

\section{Introduction}

In the study of high temperature gauge theory in thermal
equilibrium, a particularly useful approach has turned out to be
that of dimensional reduction \cite{Ginsparg:1980ef}. There, one describes the system via a
$d-1$ dimensional static effective theory built using the fact
that, at high temperature, the non-static field
modes decouple quite efficiently from the dynamics of length scales
larger than or equal to the inverse Debye mass. This has led to
several important advances, including \textit{e.g.~}new results for
the spatial string tension and various correlation lengths
\cite{ls}, as well an efficient reorganization of the weak coupling
expansion of the QCD pressure \cite{bn1}. For the latter quantity,
dimensional reduction has even provided a framework for extending
the expansion to the full $g^6$ order, which is where the pressure
obtains its first non-perturbative contributions --- a topic that has
recently received much attention \cite{klry}.

Despite the plethora of results derived utilizing dimensional reduction in perturbative setups, there is as such no reason to view the method as only a tool for weak
coupling expansions. After all, it is well-known that at least a
modest separation between the hard ($2\pi T$) and the soft and ultrasoft scales $(m_\rmi{D}, m_\rmi{mag})$
exists all the way down to the phase transition region.
A minimal extension of the current perturbative approaches would be a non-perturbative,
numerical determination of the couplings of the dimensionally reduced theory
as a function of temperature, as attempted in Ref.~\cite{Slavo}.
Another more radical --- and also more promising --- approach is to attempt to cure the lack of
center symmetry within the usual three-dimensional effective theory, EQCD,
as it is the restoration of this symmetry that drives the phase transition to the
confining phase in pure SU($N$) Yang-Mills theory.

Recently, a new effective theory for high temperature SU(3) Yang-Mills theory has been proposed that has the virtue of
respecting the Z(3) center symmetry of the original theory \cite{vy}. By construction, it cures the unphysical properties of EQCD that invalidate its use close to the deconfinement transition region, such as the fact that the physically relevant phase of EQCD is in fact only metastable \cite{klrrt}. The new theory is thus expected
to be valid over a wider temperature range than EQCD, and its properties, including the detailed structure
of its phase diagram, are currently under non-perturbative study. The lattice simulations of the new theory are, however, technically quite demanding due to the high dimensionality of the respective parameter space \cite{k}.

To test the general idea behind the construction of dimensionally reduced center symmetric theories, our aim in this paper is to perform an analysis of high temperature
SU(2) Yang-Mills theory and its corresponding Z(2) invariant effective theory. The physical setup here is highly analogous to that of the SU(3) case, but has important
simplifications due to the less complicated structure of the gauge group. Instead of having to introduce ten new heavy (and strictly speaking unphysical) degrees of
freedom to ensure that a superrenormalizable theory with the correct symmetries can be constructed, it will this time be sufficient to consider only \textit{one} extra real
scalar field, whose mass is then the only adjustable parameter in the theory. This makes a marked difference from the point of view
of the lattice simulations.

A competing --- or rather complementary --- approach to the building of center-symmetric effective theories for thermal SU($N$) gauge theory is to set the requirement of renormalizability aside, but rather concentrate on the various kinds of models one can construct for the full theory Wilson line near $T_c$ \cite{pisa1,kovner}. An important step in this direction was taken in Ref.~\cite{pisa}, where a new model was introduced that in a rough sense corresponds to a non-linear sigma model version of that of Ref.~\cite{vy}. Very recently, a numerical study of this model was completed in the case of the gauge group SU(2) \cite{dumi}, and it will be interesting to investigate, to what extent their predictions will agree with those of our new theory. This task is, however, outside the scope of the present paper.

The paper is organized as follows. In Section 2, we introduce the new effective theory, write down its Lagrangian and discuss its degrees of freedom as well as the structure of its potential. We also analyze the vacua of the theory by computing the one-loop effective potential. In Section 3, we then perform a perturbative
matching of the effective theory parameters to the full theory ones. This amounts to demanding that upon integrating out the extra heavy degree of freedom of our
theory in the high-temperature limit, we obtain the
usual EQCD Lagrangian, as well as making sure that the leading order tension and profile of the Z(2) domain wall in the effective theory agree with the corresponding
quantities in the full theory. Section 4 on the other hand contains the details and results from a non-perturbative analysis of the new theory. We map its phase diagram
in terms of the two remaining parameters, the four-dimensional gauge coupling and the ratio of the mass of the heavy scalar field and the temperature, and propose
a simple scheme to match the value of the latter to the full theory.
Finally, we summarize our findings and discuss their implications on future work within the present theory and its SU(3) counterpart in Section 5. Many computational details, such as the lattice-continuum relations for the parameters of the new theory, are presented in the Appendices.

\section{The effective theory}

Our aim in this Section is to generalize the center-symmetric effective field theory introduced for SU(3) Yang-Mills theory in Ref.~\cite{vy} to the case of the gauge group SU(2). We begin this by introducing the required degrees of freedom and the Lagrangian, and then proceed to evaluate the effective potential of the theory to one loop order. Finally, by rewriting the Lagrangian of the theory in terms of fields fluctuating around the minima of the potential, we make the connection to EQCD transparent. Our approach is somewhat different from that of Ref.~\cite{vy}, in which the effective potential turned out too complicated to evaluate. As will be seen below, having an explicit expression for $V_\rmi{eff}$ at hand makes in particular the perturbative matching of the theory to the full one intuitive and easily tractable.

\subsection{Degrees of freedom}

The degrees of freedom in any center symmetric effective theory, with which one wishes to correctly describe the physics of Yang-Mills theory close to the phase transition region, must include something reminiscent of the Wilson line of the full theory,
\begin{eqnarray}
    \Omega(\mathbf{x})
    &\equiv&
     {\cal P} \, \exp
    \bigg[i \int_0^{\beta}\!\!{\rm d\tau} \> A_0(\tau,\mathbf{x})\bigg].
\label{omega}
\end{eqnarray}
Denoting this $2\times 2$ matrix field in our theory by $\Z$, we require that the action of the theory must be invariant under the local gauge transformations
\begin{equation}
    \begin{array}{l}
    \Z(\x) \to \gauge(\x) \, \Z(\x) \, \gauge(\x)^\dagger \,,
    \\[5pt]
    \mathbf A(\x)
    \to
    \gauge(\x) \, (\mathbf A(\x) + i \mathbf\nabla) \, \gauge(\x)^\dagger \,,
    \end{array}
\end{equation}
with $\mathbf A(\x)$ denoting the spatial gluons and $\gauge(\x) \in$ SU(2),
as well as under the global Z(2) phase rotations,
\begin{align}
    \Z(\x) &\to {\rm e}^{i \pi n} \, \Z(\x) = \pm  \,\Z(\x), \kern 2cm
\label {eq:Z3trans}
\end{align}
for integer $n$.

 In analogy with Ref.~\cite{vy}, we do not wish to formulate our theory in terms of an SU(2) matrix, but instead regard the field $\Z$ as a spatially coarse-grained Wilson line operator, in terms of which a super-renormalizable effective theory with only polynomial interactions can be constructed.
This implies defining $\Z$ via the block transformation
\ba
\Z(\vec{x})=\frac{T}{V_\rmi{Block}}\int_V d^3y \,
U(\vec{x},\vec{y})\Omega(\vec{y})U(\vec{y},\vec{x}),
\ea
where the integration goes over the (somewhat arbitrary) ${\mathcal O}(T^{-3})$ volume of the block and
$U(\vec{x},\vec{y})$ is a Wilson line connecting the points $\vec{x}$
and $\vec{y}$ at constant time $\tau=0$.

The crucial difference to the SU(3) case now comes from the fact that with SU(2), the coarse-graining almost preserves the group property, as a sum of SU(2) matrices is up to an overall real constant another SU(2) matrix. Taking furthermore into account that the exponentiation of a sum of Pauli sigma matrices, or the generators of SU(2), can be written as a linear combination of the very same matrices plus the unit matrix, this implies that we may parametrize $\Z$ as
\begin{eqnarray}
    \Z = \frac{1}{2}\Big\{\heavy \openone + i \light_a \sigma_a \Big\}\equiv \frac{1}{2} \heavy \openone+i \light.
\end{eqnarray}
Here, $\heavy$ and $\light_a$, $a=1,2,3$, are real scalar fields, the $\sigma_a$ denote the Pauli matrices, and the new fields are obviously given by the formulas
\ba
\heavy&=&\tr \Z,\\
i\light&=&\Z-\fr{1}{2}\tr \Z \,\openone .
\ea

Compared to the Wilson line operator of the full theory, the field $\Z$ seems to contain one extra real degree of freedom. In analogy with Ref.~\cite{vy}, we however assume this auxiliary field to be heavy (with a mass of order $T$ that originates from the coarse-graining procedure), so that at high temperatures it decouples from the physics of length scales $\gtrsim 1/(gT)$ and can be integrated out. We will identify this heavy mode later.

\subsection{The Lagrangian}

We take the Lagrangian of our effective theory to be composed of the conventional kinetic terms for the magnetic gluons and the $\Z$ field, plus a potential $V(\Z)$,
\begin{eqnarray}
    \mathcal{L}
    &=&
    g_3^{-2}\Big\{\half \, \tr  F_{ij}^2
    + \tr\! \left(D_i \Z^{\dagger}D_i\Z\right)
    + V(\Z)\Big\}
    ,
\label{lageff2}
\end{eqnarray}
with $D_i \equiv \partial_i -i [A_i,\,\cdot\,]$
and $F_{ij} \equiv \partial_i A_j - \partial_j A_i - [A_i,A_j]$. The potential $V$ is chosen to be of the most general superrenormalizable, $Z(2)$ and gauge invariant form\footnote{We want to emphasize here that unlike in the SU(3) case, our potential contains all terms allowed by symmetry and super-renormalizability, so that this time
even the purists should not object to calling the theory a true effective theory rather than just a model.}
\ba
V(\Z)&=&b_1 \heavy^2 + b_2 \light_a^2 +c_1 \heavy^4 + c_2 \(\light_a^2\)^2 + c_3 \heavy^2 \light_a^2, \label{pot1}
\ea
where, due to the superrenormalizability of the theory, the coefficients $c_i$ do not depend on the renormalization scale of the effective theory. The mass terms $b_1$ and $b_2$, on the other hand, acquire a scale dependence at two-loop level. In the $\msbar$ scheme, their expressions read
\begin{align}
b_1(\MB)&=\frac{1}{16\pi^2}\left[48c_1^2+12c_3^2+12c_3g_3^2\right]\log\left(\frac{\Lambda_{b_1}}{\MB}\right),\label{eq:ren_b1}\\
b_2(\MB)&=\frac{1}{16\pi^2}\left[80c_2^2+4c_3^2-40c_2g_3^2\right]\log\left(\frac{\Lambda_{b_2}}{\MB}\right),\label{eq:ren_b2}
\end{align}
where $\MB$ is the renormalization scale of the effective theory and $\Lambda_{b_i}$ are constants specifying the theory. In the following, we will set $\MB=g_3^2$ and
abbreviate $b_i(g_3^2)\equiv b_i$.

We can write the potential also in the alternative form
\ba
V(\Z) &=& h_1 \tr(\Z^\dagger \Z) + h_2 (\tr \Z^\dagger \Z)^2 + g_3^2\Big\{s_1 \tr \light^2 +  s_2 \(\tr \light^2\)^2 +s_3 \heavy^4\Big\}, \label{pot}
\ea
where we, motivated by a perturbative analysis, have assumed the coupling constants $h_i$ and $s_i$ to scale as $(\gE)^0$, so that the operators become naturally divided into hard and soft subsets. The reason for this is related to the transformation properties of the operators under an SU(2)$\times$SU(2) transformation of $\Z$,
\ba
\Z &\rightarrow& \Omega_1\Z \Omega_2,\;\;\; \Omega_i \in {\rm SU(2)},
\ea
which will later be seen to translate into a shift symmetry in the light physical field of the theory, when we integrate out the auxiliary heavy scalar. The SU(2)$\times$SU(2) invariant terms in the potential are responsible for the ${\mathcal O}(T)$ mass scale of this scalar field, but do not affect the perturbative matching to EQCD at leading order. They must therefore come with sufficiently hard, or order $(\gE)^0$, couplings denoted by $h_i$. The part of the potential that does not possess this additional symmetry will, on the other hand, be found to give rise to the potential and mass scale of the Goldstone mode of the hard potential, later to be identified as the $A_0$ field of EQCD. Consequently, these terms in the potential come with ${\mathcal O}(\gE^2)$ couplings, parametrized by the constants $s_i$. Finally, relating the two sets of couplings to each other produces
\ba
b_1&=&\frac{1}{2}h_1,\;\;\;\; \label{b1}
b_2\,=\,\frac{1}{2}(h_1+g_3^2s_1), \\
c_1&=&\frac{1}{4}h_2+g_3^2s_3,\;\;\;\;
c_2\,=\,\frac{1}{4}(h_2+g_3^2s_2), \;\;\;\;
c_3\,=\,\frac{1}{2}h_2. \label{c3}
\ea
Perturbatively matched results for these parameters will be listed in Section 3.3.

\subsection{The effective potential}

Before moving on to the perturbative matching of the effective theory to the full one, we address a somewhat more general question that will turn out to be of much use in the forthcoming analysis: What is the one-loop effective potential of our theory when computed in the background of constant or sufficiently slowly varying\footnote{Implying that we may neglect the derivative terms of the background fields in the classical action when computing the one-loop functional determinants in Appendix A.} classical fields
\ba
\langle\heavy\rangle &=& \heavye , \\
\langle\light_a\rangle &=& \lighte \delta_{a,3}\, .
\ea
The choice to have a non-zero expectation value only for the $a=3$ component of the $\light$ field is totally arbitrary, but naturally allowed by gauge invariance. Likewise, we could have included a non-zero $\langle A_i^a\rangle$, but left it out for brevity, as it will not be needed in the present
context.

A straightforward calculation (the details of which are given in Appendix A) gives for the effective potential in the $R_\xi$ gauge
\ba
V_\rmi{eff}&=&\gE^{-2}\(b_1\heavye^2+b_2\lighte^2+c_1\heavye^4+c_2\lighte^4+c_3\heavye^2\lighte^2\)-\fr{|\lighte|^3}{6\pi}\(2-\xi^{3/2}\)\nn
&-&\fr{1}{12\pi}\Big\{\(b_1+b_2+6\heavye^2c_1+6\lighte^2 c_2+(\heavye^2+\lighte^2)c_3-\sqrt{\eta}\)^{3/2}\nn
&+& \(b_1+b_2+6\heavye^2c_1+6\lighte^2 c_2+(\heavye^2+\lighte^2)c_3+\sqrt{\eta}\)^{3/2}\nn
&+&2\(\xi\lighte^2+2b_2+4\lighte^2c_2+2\heavye^2 c_3\)^{3/2}\Big\} + {\mathcal O}(\gE^2), \label{effpot}
\ea
where we have denoted
\ba
\eta &=&\(b_1-b_2+6(\heavye^2 c_1-\lighte^2 c_2)-(\heavye^2-\lighte^2)c_3\)^2+16\heavye^2\lighte^2 c_3^2.
\ea
The explicit gauge parameter dependence of the result should come as no surprise, as the effective potential is not a gauge invariant quantity. Only its minima structure is of physical significance.

\subsection{Minimizing the potential}

The truncation of the loop expansion of the effective potential to one loop order implicitly assumes that we are working in the weak coupling limit, so we may consistently further simplify the result of Eq.~(\ref{effpot}) by using the identities of Eqs.~(\ref{b1})--(\ref{c3}). Looking at the potential order by order in $\gE$, we first obtain
\ba
V_\rmi{eff}&=&\fr{\gE^{-2}}{4}(\heavye^2+\lighte^2)\(2h_1+h_2(\heavye^2+\lighte^2)\)+{\mathcal O}\big((\gE)^0\big), \label{perteffpot2}
\ea
which, depending on the sign of $h_1$ (note that perturbative stability requires $h_2>0$), is either minimized by $\heavye = \lighte = 0$ or
\ba
\heavye^2+\lighte^2&=&v^2\;=\;-\fr{h_1}{h_2}+{\mathcal O}(\gE^2).
\label{effpot_min}
\ea

Upon comparison with the effective potential of the Wilson line in the four-dimensional theory, we note that the case of interest in the present paper, where we wish to be working in the high temperature (deconfined) phase of the full theory, is that of $h_1<0$. We will thus assume this to be the case from now on. Using then the fact that any ${\mathcal O}(\gE^2)$ correction to the functions $\heavye$ and $\lighte$ would only contribute to the potential at the NNLO level, we may parametrize $\heavye$ and $\lighte$ by
\ba
\heavye&=&v\cos(\pi\alpha),\\
\lighte&=&v\sin(\pi\alpha),
\ea
with $\alpha$ a real function, and insert these values into Eq.~(\ref{effpot}). This leads to
\ba
V_\rmi{eff}=\fr{s_1v^2}{2}\sin^2(\pi\alpha)+\fr{s_2v^4}{4}\sin^4(\pi\alpha)+s_3 v^4\cos^4(\pi\alpha)
-\fr{v^3}{3\pi}|\sin(\pi\alpha)|^3+{\mathcal O}(\gE^2), \label{perteffpot}
\ea
where we have dropped a trivial constant from the result.
Note that the $\xi$-dependence of the effective potential $V_\rmi{eff}$
in Eq.~(\ref{effpot}) has disappeared at the minimum of the potential, defined by
Eq.~(\ref{effpot_min}), as required by Nielsen's theorem.

The locations of the minima of the function $V_\rmi{eff}$ in Eq.~(\ref{perteffpot}) depend on the values of $v$ and $s_i$. As we want our theory to inherit the Z(2) minima structure of the effective potential of the full theory Wilson line, we will for the time being assume (and later confirm) that these parameters have values such that the potential is minimized by $\alpha = n\pi$, $n\in \mathbb{Z}$. This in turn implies that the effective potential has its minima at $\omega=0$, or
\ba
\langle \Z \rangle &=& \pm \fr{v}{2} \openone. \label{minima}
\ea

Specializing now to fluctuations around one of these physically equivalent Z(2) minima, we denote
\ba
\Z=\pm \bigg\{\frac{1}{2}v \openone + g_3\Big(\fr{1}{2}\heavys\openone + i \lights\Big)\bigg\}, \label{eq:background}
\ea
with $\lights$ a traceless hermitian matrix field, $\chi\equiv\chi_a\sigma_a/2$, and in addition rescale the three-dimensional gauge field by $A_i\to g_3 A_i$. This enables us to rewrite the Lagrangian of the effective theory in the form
\ba
    \mathcal{L}
    =  
    \; \half\, \tr F_{ij}^2
    + \half \left[ \(\partial_i \heavys\)^2 + m_0^2 \, \heavys^2 \right]
    + \tr\!\big[ (D_i \, \lights)^2 + m_\lights^2 \, \lights^2 \big]
    + V_s(\heavys,\lights) \,,
\label{lag}
\ea
where the masses of the shifted fields have become
\begin{align}
m_0^2 & \equiv m_{\heavys}^2= 8v^2 c_1 = -2 h_1, \label{mass1} \\
m_\lights^2&= 2 \(b_2 +v^2 c_3\) =\(s_1-4v^2s_3\)g_3^2.
\end{align}
The parametrically heavier field $\phi$ is hereby identified as the auxiliary degree of freedom in our theory, which is to be integrated out in the high temperature limit. Finally, given in terms of the shifted fields, the potential $V_s$ reads
\ba
V_s &=& 2g_3 v\Big\{2c_1 \heavys^3 + c_3 \heavys \lights_a^2\Big\}
+ g_3^2\Big\{c_1 \heavys^4 + c_2 \(\lights_a^2\)^2 + c_3 \heavys^2 \lights_a^2\Big\} \label{eq:Vs}\\
&=& g_3 vh_2\heavys\Big(\heavys^2 + \lights_a^2\Big)
+ \fr{g_3^2h_2}{4}\Big(\heavys^2 + \lights_a^2\Big)^2
+4g_3^3 v s_3 \heavys^3
+ \fr{g_3^4}{4}\Big\{4s_3 \heavys^4 + s_2 \(\lights_a^2\)^2\Big\}.\nonumber
\ea
The Z(2) symmetry is now realized by a simultaneous sign flip of
$v$ and $\heavys$.

\section{Parameter matching at high temperature}

In the previous Section, we saw that at least under certain assumptions on the values of its parameters, our effective theory possesses a minima structure reminiscent of that of the underlying full Yang-Mills theory. Motivated by the connection between our degree of freedom $\Z$ and the Wilson line operator $\Omega$, it is therefore tempting to associate our light field $\lights$, describing traceless anti-Hermitian fluctuations around the deconfining minima, with the $A_0$ field of EQCD. To make this correspondence more concrete, we now move on to the high-temperature limit, where we integrate the heavy fluctuation field $\heavys$ out from the effective theory and then match the Lagrangian of the resulting 3d Yang-Mills plus adjoint Higgs theory to that of EQCD. Following the approach of Ref.~\cite{vy}, we will also require that the leading order Z(2) domain wall stretching between the two deconfined minima of the effective theory has the correct tension and width dictated by the full theory. It will be seen that this is sufficient to fix all but one of the coupling constants of our theory.

\subsection{Reduction to EQCD}
The process of integrating out the heavy field $\heavys$ is greatly simplified by the SU(2)$\times$SU(2) invariance of the hard part of our potential, as it is easily seen to translate into a shift symmetry in $\lights$ (for details, see Ref.~\cite{vy}). This implies that upon integrating the heavy field out, the graphs with only $h$-vertices will have to cancel, and, in particular, that at leading order we may simply read off the parameters of the resulting Lagrangian,
\begin{eqnarray}
    \mathcal{L}_{\rm light}
    &=&
    \half \, \tr  F_{ij}^2
    + \tr\! [(D_i \, \lights)^2 + m_\lights^2 \, \lights^2 + \tilde\lambda \, \lights^4]+...,
\label{lageff3}
\end{eqnarray}
from Eq.~(\ref{lag}). Doing so, we immediately obtain
\begin{equation}
    m_\lights^2 = \(s_1-4v^2s_3\)g_3^2 +{\mathcal O}\(\gE^4\)\,, \qquad
    \tilde\lambda = 2s_2 \, g_3^4+{\mathcal O}\Big(\fr{\gE^6}{m_0^2}\Big) \,,
\label {eq:lightcoeffs}
\end{equation}
where we have used the SU(2) identity $\(\tr \lights^2\)^2=2\,\tr \lights^4$. A comparison with the EQCD Lagrangian then confirms the validity of the simple identification $\lights_a \leftrightarrow A^a_0 $ and further leads to the leading order parameter values
\ba
    \gE^2 + {\mathcal O}\Big(\fr{\gE^4}{m_0}\Big)&=&g^2T+\mathcal{O}(g^4T), \label{ge}\\
    s_1-4v^2s_3+ {\mathcal O}\(\gE^2\) &= &\fr{2T}{3} + \mathcal{O}(g^2T) \,,\label{ct1}\\
    s_2+ {\mathcal O}\Big(\fr{\gE^2}{m_0^2}\Big)&=&\fr{1}{3 \pi^2 T} +{\mathcal O}\Bigl(\frac{g^2}{T}\Bigr)
    \, ,\label{ct3}
\ea
where $g$ is the four-dimensional gauge coupling and $T$ the temperature of the full theory. It is straightforward to see that these results are consistent with the assumption we made in Sec.~2.4 about the ${\mathcal O}\big((\gE)^0\big)$ effective potential being minimized by $\pi\alpha$.

\subsection{Domain wall properties}

\begin{FIGURE}[ht]
{ \centerline{\def\epsfsize#1#2{0.55#1}
    \epsfbox{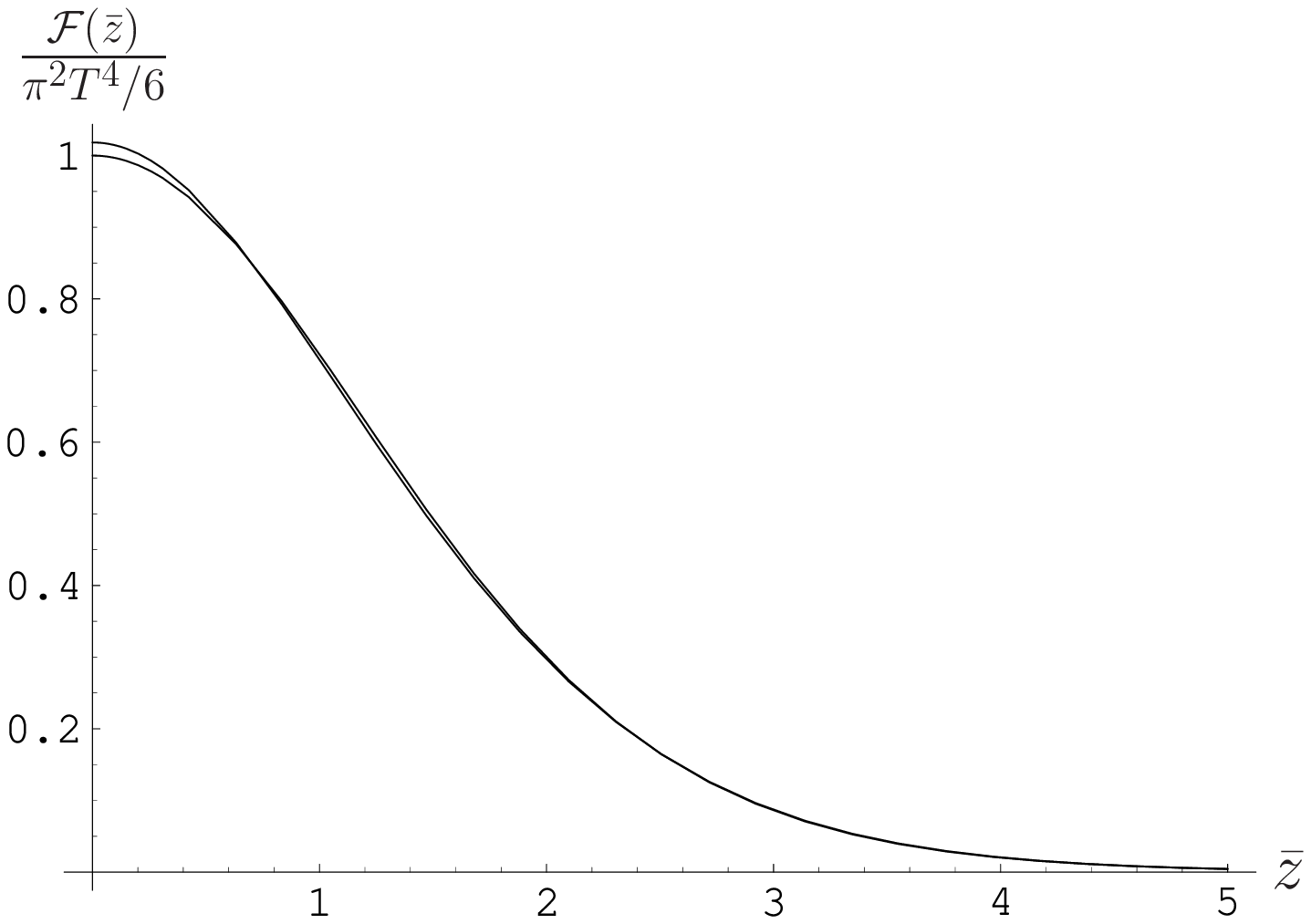}
    \epsfbox{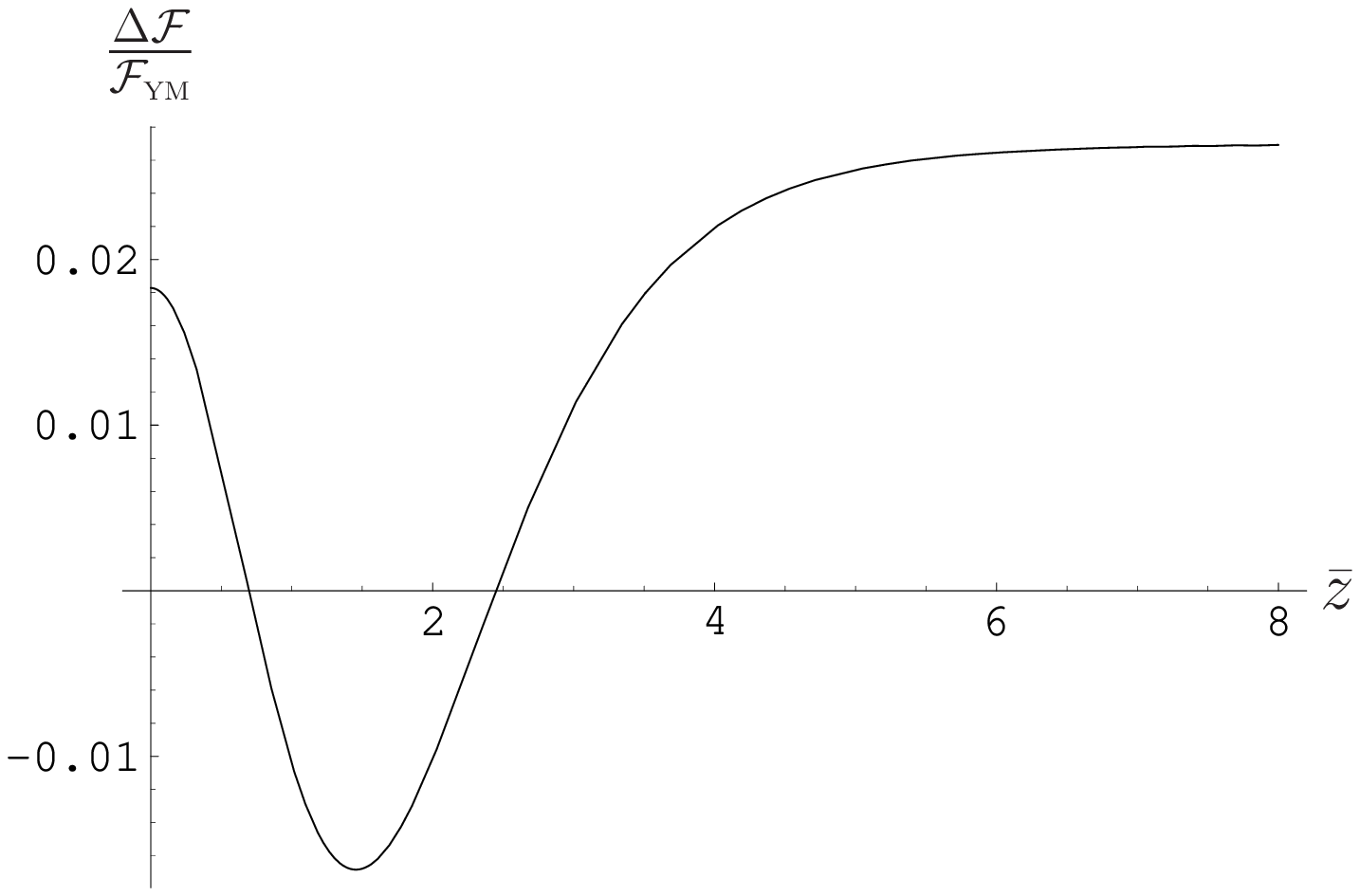}}
\caption[a] { Left: The normalized domain wall free energy densities $\mathcal F(\bar{z})$ in the effective
and full theories, plotted versus $\bar z \equiv g(T) T z$.
Right: The relative difference of the two curves,
$(\mathcal F_{\rm eff}(\bar{z})-\mathcal F_{\rm
YM}(\bar{z}))/\mathcal F_{\rm YM}(\bar{z})$. \label {fig:wall} } }
\end{FIGURE}

We now move on to the determination of the leading order Z(2) domain wall profile in the effective theory, stretching between the two physically equivalent Z(2) minima of Eq.~(\ref{minima}). It is clearly possible to have the field $\Z$ minimize the hard part of the potential throughout the wall, which implies that the width of the wall will become of order $1/(gT)$ and, in particular, that we may use the results of Sec.~2.4 for the perturbative effective potential of our theory.\footnote{The assumption we made in the derivation of Eq.~(\ref{effpot}) about the background field being \textit{sufficiently slowly varying} in order for its kinetic terms to be negligible is found to be fulfilled here, as the width of the wall being of order $1/(gT)$ implies that each spatial derivative of the background field introduces an additional factor of $g$.}
In Appendix B, we present in some detail our leading-order calculation of the
domain wall profile, including the gauge-invariant free energy density $\mathcal F(z)$,
from which we can extract the domain wall tension (by integrating
$\mathcal F(z)$ over $z$) and width (by integrating $z^2 \mathcal F(z)$).
These two quantities agree with those determined perturbatively from the
full theory at high temperature, if one chooses the parameters as
\begin{eqnarray}
    v&\approx&2.001622 T,\label{vres}\\
    s_3&\approx&1.60379/(12\pi^2T). \label{s3res}
\end{eqnarray}
We observe from Fig.~1 that with this choice, the wall profiles in the two theories are practically overlapping, with the relative error being in the per cent range.

\subsection{Leading order parameters}
To summarize our findings from the present Section, we collect here all the results from the leading order perturbative matching of the effective theory parameters. Defining the dimensionless ratio
$m_0/T = \sqrt{2h_2}\vv \equiv r$, with $\vv\equiv v/T$, we may write the couplings in terms of only $g$, $r$ and $T$, giving
\ba
b_1&=&-\frac{1}{4}r^2 T^2, \label{b1num}\\
b_2&=&-\frac{1}{4}r^2T^2+0.441841 g^2T^2,\\
c_1&=&0.0311994r^2+0.0135415g^2,\\
c_2&=&0.0311994 r^2+0.008443432g^2,\\
c_3&=&0.0623987r^2, \label{c3num}
\ea
and somewhat trivially
\ba
g_3^2 &=& g^2 T  \,. \label{gmatch}
\ea
These results will be put to use in the following Sections.

\section{Beyond perturbation theory}

Having the perturbatively matched theory at hand, an immediate question to ask becomes, what happens to its properties when one leaves the weakly coupled regime we have been investigating so far. To answer this, we now move on to consider non-perturbative numerical simulations of the theory defined by Eqs.~(\ref{lageff2}), (\ref{pot1}) and (\ref{b1num})--(\ref{gmatch}). Ideally, one would of course like to be able to analyze the theory in all generality, \textit{i.e.~}in terms of completely general couplings $b_i$, $c_i$, but as we will see in the following, the two-dimensional slice spanned by $r$ and $g$ already provides a physically motivated, interesting and, above all, numerically controllable subset of the entire parameter space.

\subsection{Numerical simulations}

In the present paper, we wish to address the physically perhaps most pertinent question of what the phase structure of our theory looks like in the $(r,g)$ plane.
From the evaluation of the one-loop effective potential, we already know that for any finite real value of $r$, the theory will always find itself in the deconfined
phase in the limit of $g\rightarrow 0$, but in order to look at non-zero values of $g$, we must formulate it on a discrete space-time lattice. This procedure consists
essentially of two parts: First, writing down the action of the theory in a discrete form such that it reproduces the same long distance physics that one would obtain in
the continuum $\msbar$ renormalization scheme and, second, performing numerical simulations.

The first goal can be achieved by naively discretizing the action and
then matching the parameters of the discretized theory to the $\msbar$ ones, imposing the condition that the pole masses and interaction vertices of the two
theories coincide. The matching can be performed to any desired order in the lattice spacing $a$, due to the superrenormalizability of the theory, within the framework
of lattice perturbation theory. Here, we perform the calculation up to
$\mathcal{O}(a)$ (or equivalently $\mathcal{O}(1/\beta)$, where $\beta = 4/(a g_3^2)$
is the lattice coupling), thus ensuring that all non-perturbative properties of the lattice
theory coincide with those of our continuum effective theory in the continuum limit $\beta\to\infty$.
For further technical details, we refer the reader to Appendix C and the corresponding calculation
performed in the SU(3) case in Ref.~\cite{k}.
Our lattice action is defined by Eqs.~(\ref{XX})--(\ref{eq:2loop_b2}).
Note the manifest Z(2) symmetry $\Sigma \leftrightarrow -\Sigma$.

\begin{FIGURE}[t]
{ \centerline{\def\epsfsize#1#2{0.6#1}
    \epsfbox{b12n64r5.eps}\;\;
    \epsfbox{b6n64r5.eps}
}
\caption[a] {The expectation value $\langle |\Sigma| \rangle$ and susceptibility $\langle |\Sigma|^2\rangle-\langle |\Sigma| \rangle ^2$ as functions of $1/g^2$ at $r^2=5$ with $(\beta,\,N)=(12,\,64)$ [left]
and  $(6,\,64)$ [right]. The critical region shrinks as the physical volume is enlarged.\label{fig:sigma} } }
\end{FIGURE}

In our simulations, we have used the Metropolis algorithm for updating the $\heavy$ and $\light$ fields, and the Kennedy-Pendleton heatbath and overrelaxation algorithms for
the link variables. At small enough $g$, the theory is seen to reside in the deconfined phase,
but for each $r$, there is some non-zero value of $g$, at which the system exhibits a Z(2)-restoring phase transition to the confined phase, characterized by
the vanishing of the expectation value $\langle\heavy\rangle$.
There is no visible discontinuity in $\langle \heavy \rangle$, which suggests a second order transition, as expected from universality arguments (Z(2) symmetry at $d=3$). For details, see Fig.~\ref{fig:sigma}.

Next, we determine the pseudo-critical point $g_c(r)$ from the peak of the susceptibility,
\ba
\chi_\rmii{$\Sigma$} &\equiv&  \langle |\Sigma|^2\rangle-\langle |\Sigma| \rangle ^2,
\label{eq:susc}
\ea
fitting the three highest points with a second-order polynomial and estimating the error
using ten jackknife blocks.
In the thermodynamical limit, $g_c(r)$ is known to coincide with the physical critical point.
To further ascertain the universality class of the transition, we measure the Binder
cumulant $B_4 = \langle\heavy^4\rangle / \langle\heavy^2\rangle^2$.
In Fig.~\ref{fig:B4}, we exhibit our measurements for several lattice sizes $N$ as functions of the
rescaled variable $(1/g^2 - 1/g^2_c) N^{1/\nu}$, where we take the $d=3$ Ising value $0.63$
for the critical exponent $\nu$~\cite{Hasenbusch:1998gh}.
A satisfactory collapse of the data is observed at $r^2=5$
and $10$ (see Fig.~\ref{fig:B4}), and the value of the cumulant at $g=g_c$ is consistent
with the Ising value 1.604.

\begin{FIGURE}[t]
{ \centerline{\def\epsfsize#1#2{0.5#1}
    \epsfbox{binder4r5.eps}\;\;\;\;\;\;
    \epsfbox{binder4r10.eps}
}
\caption[a] {A check of the universality class of the transition:
The Binder cumulant $\langle\heavy^4\rangle / \langle\heavy^2\rangle^2$
is shown as a function of the rescaled variable $(1/g^2 - 1/g^2_c) N^{1/\nu}$
for $r^2=5$ (left) and $10$ (right).
The pseudocritical coupling $g^2_c$ is the value of the coupling, which maximizes the quantity $\chi_\rmii{$\Sigma$}$
in Eq.~(\ref{eq:susc}), and $\nu=0.63$ as appropriate for a three dimensional Z(2)-transition.
A satisfactory data collapse is observed for various volumes, and the cumulant value
at $g^2_c$ is consistent with the $3d$ Z(2) value 1.604.
\label{fig:B4} } }
\end{FIGURE}

To obtain the phase diagram of the continuum theory, we must perform thermodynamical and continuum extrapolations, $Na\to\infty$ and $a\rightarrow 0$.
The leading finite-size corrections to the critical coupling $g_c$ are described by
standard renormalization group analysis, and are of the form
\ba
1/g_c^2(Na)=1/g_c^2(\infty)+C\,(Na)^{-1/\nu},
\ea
where $g_c(Na)$ is the pseudo-critical value of the coupling which maximizes the
susceptibility, Eq.~(\ref{eq:susc}), on a lattice of size $Na$, and $C$ is a constant fitted
to our data.
This ansatz describes our large-volume data well, as illustrated in Fig.~\ref{fig:thermoexp}.
Note, however, that the finite-size effects increase as $r$ is reduced,
which is no surprise, as $r$ determines the mass $m_0$ of our heavy degree of freedom,
whose Compton wavelength increases as $1/r$. This means that very large lattices
are required to study the regime $r\ll 1$.

\begin{FIGURE}[t]
{ \centerline{\def\epsfsize#1#2{0.8#1}
    \epsfbox{thermoexp2.eps}
   }
\caption[a] {The behavior of the pseudo-critical coupling $1/g_c^2$ for fixed $r^2=5$ and 10, as one increases the extent of the lattice, $N$. The solid and dotted curves are the renormalization group scaling fits to the four and three rightmost datapoints, respectively, and the two dashed horizontal lines correspond to the $N\rightarrow \infty$ limits of these curves. The effect of including or neglecting the fourth data point is of the same order of magnitude as the statistical errors in the fits. We use this to quantify the systematics of our fits, and add the two sources of error in the final error estimate of our extrapolation.\label{fig:thermoexp} } }
\end{FIGURE}

Turning then to the continuum extrapolation, Fig.~\ref{fig:contexp} shows our
measurements of $1/g^2_c$ in a small, constant physical volume $(N a(\beta))^3$
as functions of $1/\beta$.
Our data are consistent with a discretization error linear in $1/\beta$, \textit{i.e.~}proportional to the lattice spacing $a$.
The origin of this $\mathcal{O}(a)$ error
can be traced back to the imperfect matching of the parameters of the lattice
potential with their continuum counterparts, displayed in Eqs.~(\ref{XX}): An $\mathcal{O}(a)$
error remains, which would take a three-loop calculation to eliminate.
Therefore, we extrapolate our data linearly to $1/\beta=0$.
In any case, the discretization error is small, which we expect to remain
true as long as all physical length scales of our theory are sufficiently
large compared to the lattice spacing:
\ba
a&\ll&\{1/(g^2T),1/(gT),1/(rT)\}.
\ea
The last of these inequalities limits our ability to study the regime of large $r$.
Therefore, we have concentrated our simulation efforts on $r^2=5$ and $10$,
for which the thermodynamical and continuum extrapolations are both under good control.

\subsection{Phase diagram and the matching of $r$}

Having taken into account the errors caused by the finite values of the lattice size and spacing
as explained above, we obtain the phase diagram of the continuum effective theory exhibited
in Fig.~\ref{fig:phasediagram}. In addition to our two data points, we show the special point
at $r=0$, as obtained from Ref.~\cite{phi4}. While for $r=0$ our numerical accuracy is severely hampered
by problems in the thermodynamical extrapolation, such simulations are not needed, as for $r=0$ the field $\heavy$ completely decouples from the rest. The relevant part of the
action then becomes that of a three-dimensional one-component $\lambda \phi^4$ theory, the phase diagram of which has already been carefully determined in Ref.~\cite{phi4}. This reference provides us with an accurate value for the end point of the critical line, as explained below.

\begin{FIGURE}[ht]
{ \centerline{\def\epsfsize#1#2{0.65#1}
    \epsfbox{contlimit.eps}
   }
\caption[a] {The effects of the finite lattice spacing on the value of
the pseudo-critical coupling $1/g_c^2$ in a small fixed physical volume
$V=\(16/3\)^3/g_3^6$, with $(\beta,N)=(6,8),(9,12),(12,16)$, and
$(18,24)$. The effect of finite lattice spacing is estimated from the
slope of a straight fit to the data points giving
$1/g_c^2=0.287(7)+0.14(5)/\beta$, with $\chi^2/\rm{dof}=1.03/2$, for
$r^2=5$ and $1/g_c^2=0.37(1)+0.21(8)/\beta$, with
$\chi^2/\rm{dof}=0.96/2$, for $r^2=10$. \label{fig:contexp} } }
\end{FIGURE}

Quoting Ref.~\cite{phi4}, the critical mass of a theory defined by the Lagrangian
\ba
\mathcal{L}&=&\frac{1}{2}(\partial_i\phi)^2+\frac{1}{2}m^2\phi^2+\frac{\lambda}{4!}\phi^4
\ea
reads
\ba
\frac{m^2(\bar{\mu}=\lambda/3)}{\lambda^2}&=&0.0015249(48).
\ea
A conversion of this result to our theory and the renormalization scale $\bar{\mu}=g_3^2=g^2T$
then reveals that at $r=0$, the critical value of our gauge coupling $1/g^2$ reads\footnote{A consistent value $u^*_{R+}=\lambda/m \approx 24$ has also been obtained in
Ref.~\cite{Munster}.}
\ba
1/g_c^2|_{r=0}&=&0.025543. \label{phi4point}
\ea
This point fixes the form of our critical line at small $r$, and shows that there is a critical coupling of $g\approx 6.3$, beyond which the effective theory is always
in the confined phase, no matter how small one dials the mass of the heavy scalar field.

\begin{FIGURE}[ht]
{ \centerline{\def\epsfsize#1#2{0.75#1}
    \epsfbox{phasediag2.eps}
   }
\caption[a] {The phase diagram of the perturbatively matched theory on the $(r^2,1/g^2)$ plane. The solid blue data points are from our numerical simulations and the open red one has been obtained from the known location of the critical point of $\lambda\phi^4$ theory, Eq.~(\ref{phi4point}). The curve connecting the points has been obtained from a polynomial fit and has been included in the diagram to guide the eye. \label{fig:phasediagram} } }
\end{FIGURE}

Moving on to non-zero values of $r$, we observe that the confined and deconfined phases
in our theory are separated by a second order phase transition line, which goes through
the data points $(r^2,\,1/g^2)=(5,\,0.175(4))$ and $(10,\,0.206(7))$.
Fitting a polynomial trial function to these points as well as to the known intercept
of the transition line at $r=0$, we obtain the curve shown in Fig.~\ref{fig:phasediagram}.
One observes that the curve is rather flat for $r\sim{\mathcal O}(1)$ and higher.
This can be understood by noting that the phase transition is a long distance property
of the theory, while the scale $r$ originates from the microscopic coarse-graining of the
Wilson line. Thus, while $r$ is formally an undetermined parameter of our theory, we in
fact expect (and here observe) that the properties we are interested in are rather insensitive
to it.

In particular, the latter statement implies that as $r\rightarrow \infty$, the critical curve will
approach a horizontal line, as the radial fluctuations of the
field $\Z$ become frozen and $\Z$ simply reads $v/2$ times an SU(2) matrix. In this limit, the mass of the heavy field of
our effective theory becomes so high that it decouples from the
physics and can no longer affect the properties of the system, such as the critical value of $1/g^2$. However,
the potential $V(\Z)$ in Eq.~(\ref{pot}) is not O(4)-symmetric. It has two degenerate
minima at $\heavy=\pm v, \light=0$, separated by barriers of height ${\mathcal O}(g_3^2)$.
Thus, a Z(2) transition can still take place as a function of $g^2$, leading to
a deconfined phase where $\heavy=\pm v$ at high temperature.

Finally, we would like to choose a value for the heavy mass parameter $r$, so as to best
reproduce the properties of the original $(3+1)d$ SU(2) Yang-Mills theory.
Ideally, one would like to perform a non-perturbative determination of all the couplings of the
effective action, including $r$, as a function of the temperature of the $(3+1)d$ theory,
by measuring the large-distance two-point functions of the relevant fields.
As a first simple attempt in this direction, we will now retain our perturbative determination
of the other coefficients and fix $r$ by demanding that the transition temperature be the
same in the full and the effective theories.
Assuming further a one-loop running of the full theory gauge coupling $g^2(T)$,
\ba
g^2(\bar{\mu}) &=& \fr{12\pi^2}{11\,\ln(\bar\mu/\Lambda_\rmi{QCD})},
\ea
with $\bar\mu=6.74T$ \cite{huli}, and using the result $T_c / \Lambda_\rmi{QCD}\sim 1.23$ \cite{fhk},
this amounts to fixing the coupling at the transition temperature $T_c$ to be
$g^2(T_c)\approx 5.1$. From the phase transition line of Fig.~\ref{fig:phasediagram}, we then obtain
\ba
r&=&2.6(5), \label{rres}
\ea
which in this rather crude approximation scheme is our final result for the parameter.

\subsection{Discussion}

We have formulated our effective theory on a discrete three-dimensional lattice, which has enabled us to investigate its properties at finite couplings. The parameters of the theory used in the simulations were partially fixed by perturbative matching to the full theory, which left only the parameter $r$, related to the heavy mass scale of the effective theory, unfixed. It is an impressive demonstration of the power of respecting the correct symmetries that starting from such restrictive assumptions, clearly not fully consistent close to the phase transition region,
we were able to find a line of second order phase transitions in the $(r,g)$ plane, just as the full theory predicts.

The success of the new effective theory provides a stark contrast to the case of EQCD, which cannot reproduce the phase structure of the full theory, as it inherently only describes small fluctuations of the $A_0$ field around one of the Z(2) minima.
The two deconfining minima of this theory are therefore physically inequivalent and only resemble each other in the immediate vicinity of the critical point, which furthermore is located far away from the physical line of matching
\cite{Hart:1996ac}. In addition, in the absence of an order parameter, there is no clear signature of a confining phase, whereas our theory provides a consistent description of it.

Having said this, it should nevertheless be emphasized that there is still work to be done within the new effective theory. First of all, while the perturbative matching we have
performed in the present paper ensures that the theory has the correct behavior at asymptotically high temperatures, one would like to quantify the extent to which it is able to
capture the dynamics of the phase transition. To this end, one should compute various physical
quantities in the effective theory and compare them to results in the full theory as a function of the temperature; among these are \textit{e.g.~}the spatial string tension and the domain wall tension, as well as the entire spectrum of the effective theory, which is related to the screening masses of the full theory. After the matching of $r$ in Eq.~(\ref{rres}) is taken into account, our theory is in principle fully predictive, so all of these comparisons will be non-trivial tests of the validity of the assumptions and claims we have made.

While our effective theory should already at the present stage give a good approximate description of
high-temperature Yang-Mills theory down to significantly lower temperatures than EQCD, there are also various ways to improve the approximation further, such as by considering a non-perturbative
determination of the effective theory couplings. The simplest strategy would
be to keep the present perturbative matching of all coefficients, and let
$g$ and $r$ evolve non-perturbatively with $T$. The gauge coupling $g$ can be
determined from simple measurements of the spatial string tension or the
domain wall tension as already done in Ref.~\cite{dfn}. The parameter $r$, or equivalently
the mass of the heavy field which represents the magnitude of a coarse-grained
Wilson line, can on the other hand be also determined in the full theory,
by performing a blocking transformation
on the Wilson line and monitoring the correlator of its magnitude.
These two non-perturbative matchings are straightforward to implement,
and should improve the quantitative description the effective theory provides at
temperatures close to $T_c$.

\section{Conclusions}

Building a dimensionally reduced effective theory for high-temperature SU($N$) Yang-Mills theory that respects all symmetries of the latter is an important task
not only from the point of view of completeness and rigor, but also because there is genuine interest to study the extent to which the failure of EQCD to respect the
center symmetry explains its unphysical properties close to the phase transition region. To this end, the task of supplementing EQCD with just enough heavy degrees of freedom
to enable the formulation of a superrenormalizable center-symmetric theory was tackled in Ref.~\cite{vy}, where a new effective theory for SU(3) Yang-Mills theory
was built. An unfortunate practical problem, however, turned out to be the high dimensionality of the respective parameter space, as well as the fact that not all
operators allowed by the symmetries could be included in the potential for practical purposes. Consequently, numerical studies of the theory have turned out to be very impractical \cite{k}.

As conventional wisdom states that $2\ll 3$, it is not surprising that SU(2) Yang-Mills theory provides a technically much simpler platform to investigate the same issues one encounters with SU(3) Yang-Mills theory. In the present paper, we have transcribed the ideas of Ref.~\cite{vy} to the case of the simpler gauge group, and doing this have managed to include in the Lagrangian all superrenormalizable operators allowed by the symmetries. In addition, we have seen that the number of heavy scalar degrees of freedom one needs to introduce in the theory as well as the number of free parameters not fixed by simple perturbative matching are both one. This has enabled us to extend the matching to include the determination of the mass scale of our heavy field, using as matching criterion the simple requirement that the phase transition of the effective theory take place at the same temperature as that of the full one. More generally, we have been able to map the phase diagram of the new theory as a function of the mass parameter $r$ and the four-dimensional gauge coupling $g$, revealing a second order phase transition line as predicted by the full theory. This required performing non-perturbative lattice simulations, as well as a two-loop lattice perturbation theory calculation.

Our primary motivation in the present paper has been to address the question, whether the region of applicability of dimensionally reduced effective theories of hot Yang-Mills theory can be extended down to the phase transition region, provided one correctly accounts for all the symmetries of the full theory. Our findings regarding the phase structure of our new effective theory are highly encouraging in this respect:
The spurious $\langle A_0 \rangle \neq 0$ phase of EQCD disappears, and
we observe instead a genuine Z(2) finite temperature transition, which
can be adjusted to occur at the right temperature.
Nevertheless, more work is required to put our approach on solid
quantitative ground.
In this process, the framework laid out in the present paper will be an invaluable asset, as the theory is now fully predictive and simply awaits further simulations to be performed. Beyond this, obvious further challenges include the generalization of our work to the case of SU(3), as well as the introduction of soft center-symmetry breaking operators in the effective theory with the purpose of modeling the presence of dynamical quarks in the full one.

\section*{Acknowledgments}
We are grateful to Keijo Kajantie, Mikko Laine, Kari Rummukainen and Larry Yaffe for useful discussions. P.dF.~and A.V.~would like to thank the Isaac Newton Institute for Mathematical Sciences, Cambridge, and A.K.~the Institute for Nuclear Physics, Seattle, and the Institute of Theoretical Physics, TU Vienna, for hospitality. A.K.~has been supported by the Academy of
Finland, contract number 109720, the EU I3 Activity RII3-CT-2004-506078 HadronPhysics and the Jenny and Antti Wihuri Foundation, while A.V.~has been supported in part by the Austrian Science Foundation, FWF, project No.~M1006, as well as the U.S.~Department of Energy under Grant No.~DE-FG02-96ER40956. The numerical simulations were carried out at CSC - Scientific Computing Ltd., Finland.

\appendix
\section{Calculation of the effective potential}
The evaluation of the effective potential proceeds in the standard way of adding to the tree level potential of Eq.~(\ref{pot1}) the contributions from the one-loop
fluctuations of the $\heavy$, $\light$ and gluon fields. For this purpose, we write our scalar fields in the form
\ba
\heavy &=& \heavye+\gE\heavys,\\
\light_a &=& \lighte \delta_{a,3}+\gE\lights_a,
\ea
where $\heavys$ and $\lights$ will be referred to as the fluctuation fields from now on. The one-loop functional determinants are computed in the most straightforward way by choosing to work in the renormalization, or $R_\xi$ gauges, defined by the gauge fixing function
\ba
G_a&=&\frac{1}{\sqrt{\xi}}(\partial_i A_i^a+\xi g_3\omega\(\delta_{a1}\chi_2-\delta_{a2}\chi_1\)).
\ea
This enables us to decouple the gluons from the $\lights$ field at the level of the quadratic action, leaving as the only non-diagonal operator the mixed mass term of $\heavys$ and $\lights_3$.

The general structure of our one-loop potential now becomes
\ba
V_\rmi{eff}&=& \frac{1}{\gE^2}V_\rmi{cl}+ V_\rmi{A}+ V_\rmi{gh}+ V_{\heavys \lights}+V_{\lights},
\ea
where $V_\rmi{cl}$ denotes the potential of Eq.~(\ref{pot1}), evaluated with the classical fields, and the remaining terms are the contributions from the one-loop functional determinants of the quadratic gluon, ghost, $\heavys$ and $\lights$ actions, respectively. The second-to-last term $V_{\heavys \lights}$ corresponds to the contribution of the entangled $\heavys$ and $\lights_3$ fields, while $V_{\lights}$ denotes the one-loop determinant coming from $\lights_1$ and $\lights_2$.

A straightforward calculation produces for each of the above functions
\ba
V_\rmi{cl}&=& b_1\heavye^2+b_2\lighte^2+c_1\heavye^4+c_2\lighte^4+c_3\heavye^2\lighte^2,\\
V_\rmi{A} &=&\fr{1}{2}\tr\ln\bigg\{\Big(-\delta_{ij}\partial^2+\(1-1/\xi\)\partial_i\partial_j\Big)\delta^{ab} +
\lighte^2\(\delta_{a,1}\delta_{b,1}+\delta_{a,2}\delta_{b,2}\)\bigg\}\nn
&=&\int\!\fr{d^3k}{(2\pi)^3}\bigg(2\,\ln(k^2+\lighte^2)+\ln(k^2+\xi \lighte^2)\bigg)\nn
&=&-\fr{|\lighte|^3}{6\pi}\(2+\xi^{3/2}\),\\
V_\rmi{gh} &=& -\tr\ln\Big\{-\partial^2+\xi \lighte^2\(\delta_{a,1}\delta_{b,1}+\delta_{a,2}\delta_{b,2}\)\Big\}\nn
&=&\fr{|\lighte|^3}{3\pi}\xi^{3/2},
\ea
\ba
V_{\heavys\lights} &=& \fr{1}{2}\tr\ln\Big\{-\partial^2+b_1+b_2+6(\heavye^2c_1+\lighte^2 c_2)+(\heavye^2+\lighte^2)c_3-\sqrt{\eta}\Big\}\nn
&+&\fr{1}{2}\tr\ln\Big\{-\partial^2+b_1+b_2+6(\heavye^2c_1+\lighte^2 c_2)+(\heavye^2+\lighte^2)c_3+\sqrt{\eta}\Big\}\nn
&=&-\fr{1}{12\pi}\Big(\(b_1+b_2+6(\heavye^2c_1+\lighte^2 c_2) c_2+(\heavye^2+\lighte^2)c_3-\sqrt{\eta}\)^{3/2}\nn
&+& \(b_1+b_2+6(\heavye^2c_1+\lighte^2 c_2) c_2+(\heavye^2+\lighte^2)c_3+\sqrt{\eta}\)^{3/2}\Big),\\
V_\lights &=&
\tr\ln\Big\{-\partial^2+\xi\lighte^2+2b_2+4\lighte^2c_2+2\heavye^2
c_3\Big\}\nn
&=&-\fr{1}{6\pi}\(\xi\lighte^2+2b_2+4\lighte^2c_2+2\heavye^2 c_3\)^{3/2},
\ea
where we have denoted
\ba
\eta &=&\(b_1-b_2+6(\heavye^2 c_1-\lighte^2 c_2)-(\heavye^2-\lighte^2)c_3\)^2+16\heavye^2\lighte^2 c_3^2.
\ea
Adding up the various contributions now finally leads to Eq.~(\ref{effpot}).

\section{Effective theory domain wall}

In this Appendix, we compute the leading order profile of a domain wall that stretches between the two physical minima of our theory, Eq.~(\ref{minima}), our purpose being to find out, for which values of the parameters $\vv$ and $s_3 T$ the effective theory reproduces the domain wall tension and width of the full theory.

Taking the wall to lie in the $(x,y)$ plane and using the same gauge and parametrization of $\Z$ as in Sec.~2.4, we have
\ba
    \Z(z) &=& \fr{v}{2}{\rm diag} (e^{i\pi \alpha(z)}, e^{-i\pi \alpha(z)}) \,,
\ea
where $\alpha$ is this time required to satisfy the domain wall boundary conditions
\begin{eqnarray}
    \alpha(z=-\infty)&=&0 \,,\qquad \alpha(z=\infty)=1.
\label{eq:bc}
\end{eqnarray}
The wall Lagrangian is composed of the usual kinetic term for $\Z$ plus the one-loop effective potential, giving
\ba
{\mathcal L}_\rmi{wall}&=&g_3^{-2}\tr\! \left(D_i \Z^{\dagger}D_i\Z\right)+V_\rmi{eff}(\Z),
\ea
where $V_\rmi{eff}$ can be read off from Eq.~(\ref{perteffpot}). Using the results from  Section 3.1, Eqs.~(\ref{ge})--(\ref{ct3}), and defining
\ba
V_\rmi{eff}(\Z)&\equiv&\fr{\pi^2 v^2T}{2}U(\alpha),
\ea
we therefore see that the domain wall tension, \textit{i.e.~}its free energy per unit area, can be written in the form
\begin{eqnarray}
    F_{\rm dw}[\alpha]
    &\equiv&
    \fr{1}{2}\gE^{-1} \, (\pi \vv \, T)^2 \, T^{3/2} \,
    \int_{-\infty}^\infty
    d\zp \left\{ (\alpha')^2 + U(\alpha) \right\},\label{eq:Fdw}
\end{eqnarray}
with
\ba
U(\alpha)&\equiv& \fr{2}{3\pi^2}\sin^2(\pi\alpha)+\frac{\vv^2}{6\pi^4}\(1+12\pi^2Ts_3\)\sin^4(\pi\alpha) -
\fr {2\vv}{3\pi^3 }\bigl|\sin (\pi\alpha)\bigr|^3
\ea
and
\begin{eqnarray}
    \zp \equiv gT z\, ,\qquad\vv \equiv \fr{v}{T} .
\end{eqnarray}
Minimizing this functional, subject to the boundary conditions of Eq.~(\ref {eq:bc}), will yield the leading order domain wall profile and tension.

Following the steps outlined in Ref.~\cite{vy}, it is easy to find that
the classical equation of motion for $\alpha(\zp)$ can be written in the form
\begin{eqnarray}
    \int_{1/2}^{\alpha(\zp)} d\alpha \; U(\alpha)^{-1/2}
        &=&\zp \, ,
\label{eom1}
\end{eqnarray}
while the resulting domain wall tension is
\begin{eqnarray}
    \sigma
    &\equiv&
    F_{\rm dw}[\alpha]
        = \gE^{-1} \, (\pi \vv \, T)^2 \, T^{3/2}
    \int_{0}^{1}\! d\alpha \> {U(\alpha)^{1/2}} \,.
\label {eq:tension}
\end{eqnarray}
We define the width\footnote{Analogously to the Z(3) case \cite{vy}, we have observed that different definitions of the domain wall width lead to almost identical values for $\bar{v}$ and $s_3$. This can be attributed to the highly similar potentials (and subsequently similar wall profiles) in the effective and full theories, and may be taken as an \textit{a posteriori} justification of leaving out the non-superrenormalizable terms from the effective theory Lagrangian.}
$\Delta z$ of the domain wall as the square
root of the ratio of the second moment of the domain wall free energy
density and the tension, which gives
\begin {eqnarray}
    (\Delta z)^2
    &\equiv&
    \frac {(\pi \vv)^2 \, T^{5/2}} {\gE^3 \, \sigma}
    \int_{-\infty}^\infty d\zp \> \zp^2 \> U(\alpha(\zp))
\nonumber\\[5pt]
    &=&
    \frac {2(\pi \vv)^2 \, T^{5/2}} {\gE^3 \, \sigma}
    \int_{1/2}^1 d\alpha_1 \> {U(\alpha_1)^{1/2}}
    \(\int_{1/2}^{\alpha_1} d\alpha_2 \> {U(\alpha_2)}^{-1/2}\)^2
    \,.
\label {eq:width}
\end {eqnarray}
The integrations in these three equations are trivial to perform numerically for different values of the unknown parameters $\vv$ and $s_3 T$.

In high temperature $SU(2)$ Yang-Mills theory, the domain wall free energy density has been found to equal \cite{kap}
\begin{eqnarray}
    \mathcal F_{\rm YM}(z)&=& \fr{\pi^2 \, T^4}{6\cosh^{4}(\zp/\sqrt{6})} \,,
\label{eq:FYM}
\end{eqnarray}
resulting in the domain wall tension
\begin{eqnarray}
    \sigma_\rmi{YM}
    &=&
    \int_{-\infty}^\infty dz \> \mathcal F_{\rm YM}(z)
    =  \(\fr{2}{3}\)^{3/2}\fr{\pi^2T^3}{g(T)}\,
\end{eqnarray}
and width squared
\begin{equation}
    (\Delta z_{\rm YM})^2
    = 
    \sigma_{\rm YM}^{-1}
    \int_{-\infty}^\infty dz \> z^2 \, \mathcal F_{\rm YM}(z)
    = 
  \fr{\pi^2/2 - 3}{g^2(T) \, T^2} .
\end{equation}
A straightforward numerical calculation shows that the effective theory
reproduces the latter two results with the parameter values given in Eqs.~(\ref{vres})--(\ref{s3res}). Using these values, the energy density profiles of the wall in the full and effective theories are found to agree within a few per cent, as demonstrated in Fig.~1.

\section{Setup for the numerical simulations}

Using standard Wilson discretization and denoting lattice quantities with hats, the lattice action corresponding to our effective theory reads
\begin{align}
S_a &= S_W + S_\Z + V(\hat{\heavy},\hat{\light}), \label{XX}\\
S_W &=  \beta \sum_{x,i<j} \left[ 1 - \frac{1}{2} \tr[ U_{ij}] \right],\\
S_\Z &=  2 \(\frac{4}{\beta}\) \sum_{x,i}\tr \left[\hat{\light}^2 -
\hat{\light}(x)U_i(x)\hat{\light}(x+\hat{i})U^\dagger_i(x)\right] \nn
&\;\;\;+  \(\frac{4}{\beta}\)\sum_{x,i}\left(\hat{\heavy}^2(x)- \hat{\heavy}(x)\hat{\heavy}(x+\hat{i})\right),\\
V &= \(\frac{4}{\beta}\)^3 \sum_{x} \left[ \hat{b}_1 \hat{\heavy}^2 + \hat{b}_2 \hat{\light}_a^2
+ \hat{c}_1 \hat{\heavy}^4 + \hat{c}_2 \(\hat{\light}_a^2\)^2 + \hat{c}_3 \hat{\heavy}^2 \hat{\light}_a^2 \right],
\end{align}
where $\beta$ is the lattice coupling constant
\ba
\beta &=& \frac{4}{a g_3^2}
\ea
corresponding to a lattice spacing $a$.

The superrenormalizability of the theory allows the matching of the fields and coefficients of the
lattice and continuum theories to be performed exactly up to the desired order in $a$ using
lattice perturbation theory. To $\mathcal{O}(a^0)$, which is needed for performing simulations with physical $\msbar$
parameters, the lattice-continuum relations read
\begin{eqnarray}
\heavy = g_3\hat{\heavy},\,\,\,\,\,
\light = g_3\hat{\light}, \,\,\,\,\,
c_i &= \hat{c}_i,
\label{eq:tree}
\end{eqnarray}
and
\ba
\hat{b}_1 &= &\;b_1/g_3^4-\frac{1}{4\pi}(6 \hat{c}_1 + 3\hat{c}_3)\frac{\Sigma}{g_3^2 a} \nn
&+& \frac{1}{16\pi^2}\left\{(48\hat{c}_1^2+12\hat{c}_3^2-12\hat{c}_3)\left[\log\frac{6}{a\MB} + \zeta\right]
-12\hat{c}_3\left[  \frac{\Sigma^2}{4} -\delta\right]\right\}+\mathcal{O}(a)\label{eq:2loop_b1}\\
\hat{b}_2 &= & \; b_2/g_3^4-\frac{1}{4\pi}(10\hat{c}_2+\hat{c}_3+2)\frac{\Sigma}{g_3^2a},\nn
&+&\frac{1}{16\pi^2}\bigg\{(80\hat{c}_2^2 + 4\hat{c}_3^2-40\hat{c}_2)\left[\log\frac{6}{a\MB}+\zeta\right]-40\hat{c}_2\left[\frac{\Sigma^2}{4}-\delta\right]\nn
&-&\left[\frac{5}{2}\Sigma^2+\frac{2}{3}\pi\Sigma-16(\delta+\rho)+8\kappa_1-4\kappa_4\right]\bigg\}+\mathcal{O}(a),
\label{eq:2loop_b2}
\ea
where Eq.~(\ref{eq:tree}) follows from tree level scaling and  Eqs.~(\ref{eq:2loop_b1}) and (\ref{eq:2loop_b2}) from two-loop calculations, and in the latter
two the numerical coefficient $\Sigma$ has no relation to the corresponding field in our effective theory.
For details of a corresponding calculation with the gauge group SU(3) and the definition of the lattice constants, see
Refs.~\cite{k} and \cite{Laine:1997dy}.
Finally, plugging in the numerical constants and setting $\MB=g_3^2$, we get
\begin{align}
  \hat{b}_1 = &\;b_1/g_3^4-\frac{2.38193365}{4\pi}(2\hat{c}_1  + \hat{c}_3)\beta \nn
&+  \frac{1}{16\pi^2}\left\{(48\hat{c}_1^2+12\hat{c}_3^2-12\hat{c}_3)\left[\log1.5\beta + 0.08849 \right] -6.9537\,\hat{c}_3\right\}+\mathcal{O}(\beta^{-1})),\label{eq:2loop_b1n}\\
\hat{b}_2 = & \;b_2/g_3^4-\frac{0.7939779}{4\pi}(10\hat{c}_2+\hat{c}_3+2)\beta\label{eq:2loop_b2n}\\
&+\frac{1}{16\pi^2}\bigg\{(80\hat{c}_2^2 + 4\hat{c}_3^2-40\hat{c}_2)\left[\log1.5\beta+0.08849 \right]-23.17895\,\hat{c}_2
-8.66687\bigg\}+ \mathcal{O}(\beta^{-1}).\nonumber
\end{align}



\begin{thebibliography}{99}




\bibitem{Ginsparg:1980ef}
  P.~H.~Ginsparg,
  Nucl.\ Phys.\  B {\bf 170} (1980) 388;
  T.~Appelquist and R.~D.~Pisarski,
  Phys.\ Rev.\  D {\bf 23} (1981) 2305.


\bibitem{ls}
 M.~Laine and Y.~Schr\"oder,
  JHEP {\bf 0503} (2005) 067
  [hep-ph/0503061];
  A.~Hart, M.~Laine and O.~Philipsen,
  Nucl.\ Phys.\ B {\bf 586} (2000) 443
  [hep-ph/0004060];
  M.~Laine and M.~Veps\"al\"ainen,
  JHEP {\bf 0402} (2004) 004
  [hep-ph/0311268];
  M.~Veps\"al\"ainen,
  JHEP {\bf 0703} (2007) 022
  [hep-ph/0701250].


\bibitem{bn1}
E.~Braaten and A.~Nieto,
Phys.\ Rev.\ D {\bf 53} (1996) 3421
[hep-ph/9510408].

\bibitem{klry}
K.~Kajantie, M.~Laine, K.~Rummukainen and Y.~Schr\"oder,
Phys.\ Rev.\ D {\bf 67} (2003) 105008 [hep-ph/0211321];
  A.~Vuorinen,
  Phys.\ Rev.\  D {\bf 68}, 054017 (2003)
  [hep-ph/0305183];
  F.~Di Renzo, M.~Laine, V.~Miccio, Y.~Schr\"oder and C.~Torrero,
  JHEP {\bf 0607} (2006) 026
  [hep-ph/0605042];
  K.~Kajantie, M.~Laine, K.~Rummukainen and Y.~Schr\"oder,
  JHEP {\bf 0304}, 036 (2003)
  [hep-ph/0304048];
  A.~Gynther, M.~Laine, Y.~Schr\"oder, C.~Torrero and A.~Vuorinen,
  JHEP {\bf 0704}, 094 (2007)
  [hep-ph/0703307];
  A.~Hietanen, K.~Kajantie, M.~Laine, K.~Rummukainen and Y.~Schr\"oder,
  JHEP {\bf 0501}, 013 (2005)
  [hep-lat/0412008];
  A.~Hietanen and A.~Kurkela,
  JHEP {\bf 0611} (2006) 060
  [hep-lat/0609015].

\bibitem{Slavo}
  S.~Kratochvila and P.~de Forcrand,
  Nucl.\ Phys.\ Proc.\ Suppl.\  {\bf 106} (2002) 522
  [hep-lat/0110138].


\bibitem{vy}
  A.~Vuorinen and L.~G.~Yaffe,
  Phys.\ Rev.\  D {\bf 74} (2006) 025011
  [hep-ph/0604100];
  A.~Vuorinen,
  Nucl.\ Phys.\  A {\bf 785} (2007) 190
  [hep-ph/0608162].


\bibitem{klrrt}
K.~Kajantie, M.~Laine, A.~Rajantie, K.~Rummukainen and M.~Tsypin,
JHEP {\bf 9811} (1998) 011
[hep-lat/9811004].


\bibitem{k}
  A.~Kurkela,
  Phys.\ Rev.\  D {\bf 76} (2007) 094507
  [0704.1416 [hep-lat]];
 A.~Kurkela,
  PoS {\bf LAT2007} (2007) 199
  [0711.1796 [hep-lat]].


\bibitem{pisa1}
  R.~D.~Pisarski,
  Phys.\ Rev.\  D {\bf 62} (2000) 111501
  [hep-ph/0006205].


\bibitem{kovner}
A.~Kovner,
[hep-ph/0009138];
P.~Bialas, A.~Morel and B.~Petersson,
Nucl.\ Phys.\ B {\bf 704} (2005) 208
[hep-lat/0403027];
K.~Holland and U.~J.~Wiese,
[hep-ph/0011193].


\bibitem{pisa}
  R.~D.~Pisarski,
  Phys.\ Rev.\  D {\bf 74} (2006) 121703
  [hep-ph/0608242].



\bibitem{dumi}
  A.~Dumitru and D.~Smith,
  0711.0868 [hep-lat].


\bibitem{Hasenbusch:1998gh}
  A.~Pelissetto and E.~Vicari,
  Phys.Rept. 368 (2002) 549-727 [cond-mat/0012164].

\bibitem{phi4}
  X.~p.~Sun,
  Phys.\ Rev.\  E {\bf 67} (2003) 066702
  [hep-lat/0209144].

\bibitem{Munster}
  C.~Gutsfeld, J.~K\"uster and G.~M\"unster,
  Nucl.\ Phys.\  B {\bf 479} (1996) 654
  [cond-mat/9606091].

\bibitem{huli}
  S.~z.~Huang and M.~Lissia,
  Nucl.\ Phys.\  B {\bf 438} (1995) 54
  [hep-ph/9411293];
K.~Kajantie, M.~Laine, K.~Rummukainen and M.~E.~Shaposhnikov,
Nucl.\ Phys.\ B {\bf 503} (1997) 357
[hep-ph/9704416].




\bibitem{fhk}
  J.~Fingberg, U.~M.~Heller and F.~Karsch,
  Nucl.\ Phys.\  B {\bf 392} (1993) 493
  [hep-lat/9208012].

\bibitem{Hart:1996ac}
  A.~Hart, O.~Philipsen, J.~D.~Stack and M.~Teper,
  Phys.\ Lett.\  B {\bf 396} (1997) 217
  [hep-lat/9612021].


\bibitem{kap}
T.~Bhattacharya, A.~Gocksch, C.~Korthals Altes and R.~D.~Pisarski,
Phys.\ Rev.\ Lett.\  {\bf 66} (1991) 998;

\bibitem{dfn}
  P.~de Forcrand and D.~Noth,
  Phys.\ Rev.\  D {\bf 72} (2005) 114501
  [hep-lat/0506005].


\bibitem{Laine:1997dy}
  M.~Laine and A.~Rajantie,
  Nucl.\ Phys.\  B {\bf 513} (1998) 471
  [hep-lat/9705003].




\end{thebibliography}
\end{document}